\documentclass[twocolumn, superscriptaddress, preprintnumbers,amsmath,amssymb]{revtex4}

\usepackage{booktabs}
\usepackage{color,graphicx}
\usepackage{amsmath}
\usepackage{dcolumn}
\usepackage{bm}                  
\usepackage{soul}
\usepackage{enumerate}
\usepackage{multirow}
\usepackage{mathrsfs}
\usepackage{braket}
\usepackage{longtable}
\usepackage{pdflscape}

\usepackage[dvipdfm,
            pdfstartview=Fit,
            CJKbookmarks=true,
            bookmarksnumbered=true,
            bookmarksopen=true,
            colorlinks,
            pdfborder=001,
            linkcolor=blue,
            anchorcolor=blue,
            citecolor=blue
            ]{hyperref}

\usepackage{float}

\begin{document}

\title{Collapse of the $N=28$ shell closure in the newly discovered $^{39}$Na and the development of deformed halos towards the neutron dripline}

\author{K. Y. Zhang}
\affiliation{Institute of Nuclear Physics and Chemistry, CAEP, Mianyang, Sichuan 621900, China}
\affiliation{State Key Laboratory of Nuclear Physics and Technology, School of Physics, Peking University, Beijing 100871, China}

\author{P. Papakonstantinou}
\affiliation{Rare Isotope Science Project, Institute for Basic Science, Daejeon 34000, Korea}

\author{M.-H. Mun}
\affiliation{Department of Physics and Origin of Matter and Evolution of Galaxy Institute, Soongsil University, Seoul 06978, Korea}

\author{Y. Kim}
\affiliation{Center for Exotic Nuclear Studies, Institute for Basic Science, Daejeon 34126, Korea}

\author{H. Yan}
\affiliation{Institute of Nuclear Physics and Chemistry, CAEP, Mianyang, Sichuan 621900, China}
\affiliation{Key Laboratory of Neutron Physics, Institute of Nuclear Physics and Chemistry, CAEP, Mianyang, Sichuan 621900, China}

\author{X.-X. Sun} \email{sunxiangxiang@ucas.ac.cn}
\affiliation{School of Nuclear Science and Technology, University of Chinese Academy of Sciences, Beijing 100049, China}
\affiliation{CAS Key Laboratory of Theoretical Physics, Institute of Theoretical Physics, Chinese Academy of Sciences, Beijing 100190, China}

\begin{abstract}
  Halos and changes of nuclear magicities have been extensively investigated in exotic nuclei during past decades.
  The newly discovered $^{39}$Na with the neutron number $N=28$ provides a new platform to explore such novel phenomena near the neutron dripline of the sodium isotopic chain.
  We study the shell property and the possible halo structure in $^{39}$Na within the deformed relativistic Hartree-Bogoliubov theory in continuum.
  It is found that the lowering of $2p$ orbitals in the spherical limit results in the collapse of the $N=28$ shell closure in $^{39}$Na, and a well deformed ground state is established.
  The pairing correlations and the mixing of $pf$ components driven by deformation lead to the occupation of weakly bound or continuum $p$-wave neutron orbitals.
  An oblate halo is therefore formed around the prolate core in $^{39,41}$Na, making $^{39}$Na a single nucleus with the coexistence of several exotic structures, including the quenched $N=28$ shell closure, Borromean structure, deformed halo, and shape decoupling.
  The microscopic mechanisms behind the shape decoupling phenomenon and the development of halos towards dripline are revealed.
\end{abstract}

\date{\today}

\maketitle


Atomic nuclei cannot be made from arbitrary numbers of protons and neutrons.
Their existence ends at the dripline, which marks a boundary of the nuclear territory.
Mapping the dripline has always been a major goal of modern nuclear physics, as it is crucial for understanding the nuclear force and exploring the origin of elements~\cite{Thoennessen2016Book}.
The proton dripline has been experimentally delineated up to neptunium (atomic number $Z=93$)~\cite{Zhang2019PRL}.
The neutron dripline, however, is only known up to neon ($Z=10$)~\cite{Ahn2019PRL}, because the production cross sections for the most neutron-rich isotopes are extremely low.

Recently, further efforts were dedicated to probe the neutron-rich limits beyond $Z=10$ and nine events of sodium-39 ($^{39}$Na, $Z=11$) were observed, while the existence of heavier sodium isotopes was not excluded, thus not determining its neutron dripline~\cite{Ahn2022PRL}.
The nucleus $^{39}$Na has a neutron number of $N=28$, which is normally a magic number.
The disappearance of traditional magic numbers and the appearance of new ones in exotic nuclei near the dripline have attracted a lot of attention~\cite{Ozawa2000PRL,Kanungo2002PLB,Liddick2004PRL,Becheva2006PRL,Bastin2007PRL,Hoffman2008PRL,Doornenbal2013PRL,Steppenbeck2013Nature}.
It would be therefore interesting to investigate the shell property of $^{39}$Na.

\begin{figure*}[htbp]
  \centering
  \includegraphics[width=0.8\textwidth]{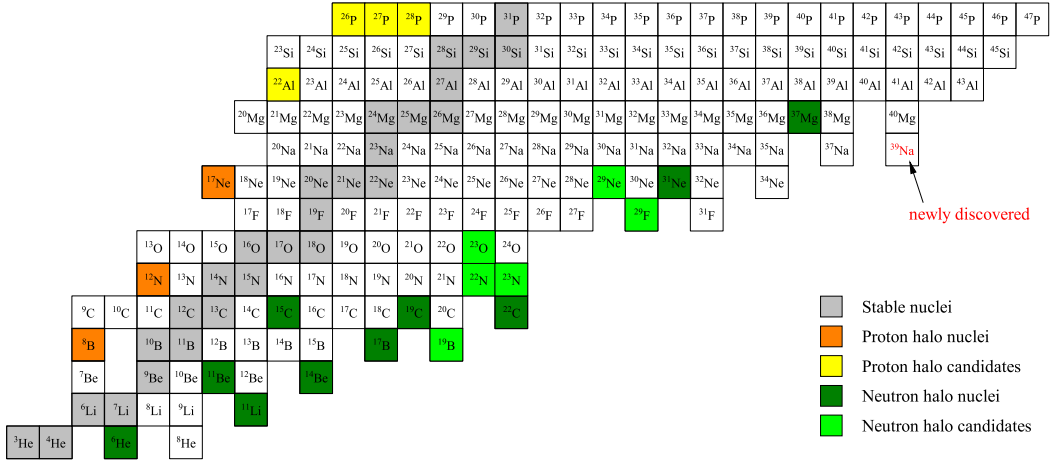}
  \caption{Experimentally known nuclear landscape from helium to phosphorus, where stable nuclei and experimentally confirmed/suggested neutron~\cite{Tanihata1985PLB,Tanihata1985PRL,Tanihata1988PLB,Kelley1995PRL,Suzuki2002PRL,Cook2020PRL,Bazin1995PRL,Bazin1998PRC,Tanaka2010PRL,Rodriguez-Tajes2011PRC,Ozawa2001NPA,Cortina-Gil2004PRL,Bagchi2020PRL,Takechi2012PLB,Nakamura2009PRL,Kobayashi2014PRL} as well as proton~\cite{Kelley1996PRL,Warner1998NPA,Kanungo2003PLB,Lee2020PRL,Navin1998PRL} halo nuclei/candidates are indicated by grey, olive/green, and orange/yellow colors, respectively. Here a confirmed halo nucleus means that both enhancement of the cross section and narrow momentum distribution are observed~\cite{Tanihata2013PPNP}. Otherwise it is treated as a candidate.}
\label{fig1}
\end{figure*}

Another novel phenomenon discovered in exotic nuclei is the halo~\cite{Tanihata1985PRL}, which also shares common interests in atomic and molecular physics~\cite{Jensen2004RMP}.
Nuclear halo phenomenon is not rare in light nuclei near the dripline.
As shown in Fig.~\ref{fig1}, it has been experimentally suggested in every isotopic chain from helium ($Z=2$) to phosphorus ($Z=15$) either on the neutron-rich or the proton-rich side~\cite{Tanihata1985PLB,Tanihata1985PRL,Tanihata1988PLB,Kelley1995PRL,Suzuki2002PRL,Cook2020PRL,Bazin1995PRL,Bazin1998PRC,Tanaka2010PRL,Rodriguez-Tajes2011PRC,Ozawa2001NPA,Cortina-Gil2004PRL,Bagchi2020PRL,Takechi2012PLB,Nakamura2009PRL,Kobayashi2014PRL,Kelley1996PRL,Warner1998NPA,Kanungo2003PLB,Lee2020PRL,Navin1998PRL}, except sodium and silicon.
The newly discovered $^{39}$Na and the worldwide development of radioactive ion beam facilities are providing an excellent platform to study more halo nuclei and, in particular, to explore whether halos exist in neutron-rich sodium isotopes.
Meanwhile, timely theoretical studies based on advanced nuclear models are desired to guide the forthcoming experiments.

In this work, we study the shell property of $^{39}$Na and explore possible halo structures in neutron-rich sodium isotopes by using the deformed relativistic Hartree-Bogoliubov theory in continuum (DRHBc)~\cite{Zhou2010PRC(R),Li2012PRC,Li2012CPL,Chen2012PRC,Zhang2020PRC,Pan2022PRC}.
Many successful applications of the DRHBc theory have been realized in the past dozen years, such as the study of halo nuclei $^{17,19}$B~\cite{Yang2021PRL,Sun2021PRC(1)}, $^{15,19,22}$C~\cite{Sun2018PLB,Sun2020NPA}, $^{31}$Ne~\cite{Zhong2022SciChina}, and $^{42,44}$Mg~\cite{Zhou2010PRC(R),Li2012PRC,Zhang2019PRC}, the investigation of deformation effects on the location of dripline~\cite{In2021IJMPE}, the prediction of stability peninsulas beyond the primary neutron dripline~\cite{Zhang2021PRC(L),Pan2021PRC,He2021CPC}, the revelation of shape coexistence from light to heavy nuclei~\cite{In2020JKPS,Choi2022PRC,Kim2022PRC}, and the exploration of rotational excitations of exotic nuclei through the combination with the angular momentum projection~\cite{Sun2021SciBull,Sun2021PRC(2)}.


The details of the DRHBc theory can be found in Refs.~\cite{Li2012PRC,Zhang2020PRC}.
Here the formalism is only briefly presented.
In the DRHBc theory, the relativistic Hartree-Bogoliubov (RHB) equation reads~\cite{Kucharek1991ZPA}
\begin{equation}\label{RHB}
\left(\begin{matrix}
h_D-\lambda & \Delta \\
-\Delta^* &-h_D^*+\lambda
\end{matrix}\right)\left(\begin{matrix}
U_k\\
V_k
\end{matrix}\right)=E_k\left(\begin{matrix}
U_k\\
V_k
\end{matrix}\right),
\end{equation}
where $\lambda$ is the Fermi energy, and $E_k$ and $(U_k, V_k)^{\rm T}$ are the quasiparticle energy and wave function, respectively.
$h_D$ is the Dirac Hamiltonian,
\begin{equation}\label{diracH}
h_D(\bm{r})=\bm{\alpha}\cdot\bm{p}+V(\bm{r})+\beta[M+S(\bm{r})],
\end{equation}
where $S(\bm{r})$ and $V(\bm{r})$ are the scalar and vector potentials, respectively.
$\Delta$ is the pairing potential,
\begin{equation}\label{Delta}
\Delta(\bm r_1,\bm r_2) = V^{\mathrm{pp}}(\bm r_1,\bm r_2)\kappa(\bm r_1,\bm r_2),
\end{equation}
with a density-dependent force of zero range,
\begin{equation}\label{pair}
V^{\mathrm{pp}}(\bm r_1,\bm r_2)= V_0 \frac{1}{2}(1-P^\sigma)\delta(\bm r_1-\bm r_2)\left\{1-\eta\left[\frac{\rho(\bm r_1)}{\rho_{\mathrm{sat}}}\right]^\gamma\right\},
\end{equation}
and the pairing tensor $\kappa$~\cite{Peter1980Book}.
In the pairing channel, $\frac{1}{2}(1-P^\sigma)$ is the projector for the spin-zero component, $\eta = 0$ corresponds to a volume pairing, $\eta=1$ $\&$ $\gamma=1$ corresponds to a surface pairing, and $\eta=0.5$ $\&$ $\gamma=1$ corresponds to a mixed one.
A finite-range pairing force, such as the Gogny~\cite{Meng1998NPA} or separable one~\cite{Tian2009PLB}, is also expected to be implemented in the future.
The pairing tensor and various densities and potentials in coordinate space are expanded in terms of the Legendre polynomials,
\begin{equation}\label{legendre}
f(\bm r)=\sum_\lambda f_\lambda(r)P_\lambda(\cos\theta),~~\lambda=0,2,4,\cdots
\end{equation}
The RHB equations~(\ref{RHB}) are solved using a Dirac Woods-Saxon basis~\cite{Zhou2003PRC,Zhang2022PRC}, which has a wave function with a more appropriate asymptotic behavior compared to the commonly used harmonic oscillator basis and is suitable for the description of weakly bound nuclei.
In Eq.~(\ref{pair}), $\eta =1$ and $\gamma=1$, i.e., a surface pairing is adopted, which has been one of the usual choices in the study of nuclear halos~\cite{Meng1996PRL,Meng1998PRL,Meng1998NPA,Meng1998PRC,Zhou2010PRC(R),Li2012PRC,Sun2018PLB,Zhang2019PRC,Sun2020NPA,Yang2021PRL,Sun2021PRC(1),Zhong2022SciChina}.
The pairing strength $V_0=-325~\mathrm{MeV~fm}^3$, the saturation density $\rho_{\mathrm{sat}}=0.152~\mathrm{fm}^{-3}$, and a pairing window of $100$ MeV are taken.
This pairing reproduces well the odd-even mass differences for not only calcium isotopes in the same mass region of $^{39}$Na, but also lead isotopes in the heavy mass region, and, thus, is suggested for the DRHBc mass table calculations~\cite{Zhang2020PRC}.
For the Dirac Woods-Saxon basis, the energy cutoff $E^+_{\mathrm{cut}}=300$ MeV and the angular momentum cutoff $J_{\max}=23/2~\hbar$ are adopted.
In Eq.~(\ref{legendre}), the Legendre expansion truncation is chosen as $\lambda_{\max}=6$.
The blocking effects in odd-mass or odd-odd Na isotopes are taken into account in the present DRHBc calculations via the equal filling approximation~\cite{Perez-Martin2008PRC,Li2012CPL,Pan2022PRC}.
The above numerical details are the same as those used in the global DRHBc mass table calculations over the nuclear chart~\cite{Zhang2020PRC,Zhang2022ADNDT,Pan2022PRC}.


Our calculations are carried out with density functionals PC-PK1~\cite{Zhao2010PRC}, PC-F1~\cite{Burvenich2002PRC}, NL3*~\cite{Lalazissis2009PLB}, NL-SH~\cite{Sharma1993PLB}, and PK1~\cite{Long2004PRC}.
In the calculated results, the binding energy of $^{38}$Na is smaller than that of $^{37}$Na, meaning that it is unstable against one-neutron emission.
$^{39}$Na is deformed in its ground state with a prolate deformation $\beta_2 \approx 0.45$ and weakly bound with a two-neutron separation energy $S_{2n} \lesssim 1$~MeV.
The results are therefore in agreement with the experimental hints that $^{39}$Na has a Borromean structure~\cite{Ahn2022PRL,Ahn2019PRL}, i.e., it is a bound three-body ($^{37}\mathrm{Na}+n+n$) system, even though no pair of its constituents is a bound system.
In the PC-PK1 results, $^{41}$Na is less deformed with $\beta_2 = 0.37$ and the last bound Na isotope, i.e., the neutron dripline, with $S_{2n} = 0.49$~MeV.
The NL3* and PK1 results also suggest $^{41}$Na as the neutron dripline, while PC-F1 and NL-SH support $^{39}$Na.
The accurate prediction of the dripline location is an ambitious goal in both nonrelativistic and relativistic density functional theories~\cite{Erler2012Nature,Afanasjev2013PLB,Xia2018ADNDT,Zhang2022ADNDT}, and it depends on the employed density functional and pairing interaction, which will not be discussed in detail in this work.
Future experimental determination of the neutron drip line as well as measurement on, e.g., quadrupole deformations along the dripline, could be helpful in constraining and optimizing the density functionals and pairing interactions.
Since PC-PK1 has been successfully used for a global description of nuclear ground-state properties~\cite{Xia2018ADNDT,Zhang2022ADNDT}, in the following we explore the possible exotic structure of $^{39}$Na based on the PC-PK1 results.

\begin{figure}[htbp]
  \centering
  \includegraphics[width=0.5\textwidth]{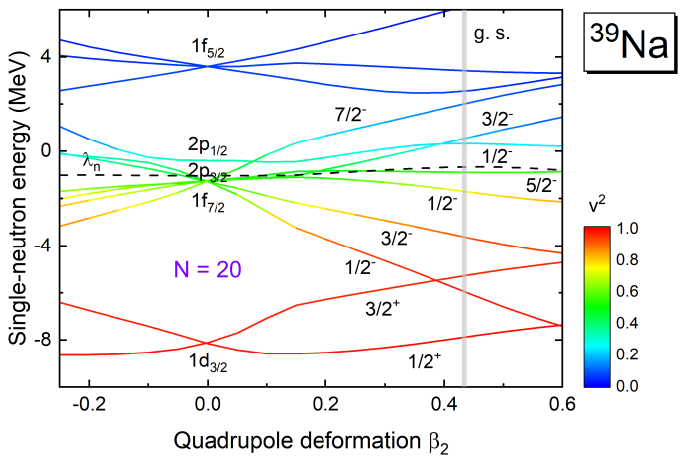}
  \caption{Single-neutron levels around the Fermi energy $\lambda_n$ (dashed line) of $^{39}$Na in the canonical basis from constrained calculations. Their quantum numbers $nlj$ in the spherical limit and $\Omega^\pi$ on the prolate side are labeled, where $\pi$ is the parity and $\Omega$ the projection of angular momentum on the symmetry axis. The occupation probability $v^2$ is scaled by colors. The grey vertical line corresponds to the ground state (g. s.) of $^{39}$Na.}
\label{fig2}
\end{figure}

The large prolate deformation of $^{39}$Na implies the collapse of the $N=28$ shell closure.
To study the shell structure of $^{39}$Na, the evolution of single-neutron levels around the Fermi energy with the quadrupole deformation obtained from the constrained calculations is shown in Fig.~\ref{fig2}.
In the spherical limit, the orbitals $2p_{3/2}$ and $1f_{7/2}$ are nearly degenerate and close to the Fermi energy, while $2p_{1/2}$ is just above them within 1~MeV and close to the particle emission threshold.
In the traditional shell model~\cite{Peter1980Book}, there is a considerable energy gap between $1f_{7/2}$ and $2p_{3/2}$, forming the $N=28$ shell closure and making the spherical shape energetically favored.
From Fig.~\ref{fig2} it becomes clear that the lowering of $2p$ orbitals results in the collapse of the $N=28$ shell closure in $^{39}$Na.
The stable quadrupole deformation of $^{39}$Na could also be explained as a result of the nuclear Jahn-Teller effect~\cite{Reinhard1984NPA}, induced by the near-degeneracy of the $pf$ orbitals in close proximity to the particle emission threshold.

Such near-degeneracy has also been predicted in $^{39}$Na and $^{40}$Mg by relativistic continuum Hartree-Bogoliubov theory with NL-SH density functional~\cite{Meng1998PLB} and the DRHBc theory with PC-F1~\cite{Sun2021PRC(2)}, respectively.
In the relativistic Hartree-Bogoliubov calculations with DD-PC1 density functional~\cite{Li2011PRC}, the single-neutron energy difference between $1f_{7/2}$ and $2p_{3/2}$ orbitals in the spherical limit is around 2 MeV for $^{40}$Mg and around 3 MeV for $^{42}$Si.
In the relativistic Hartree-Fock calculations with DD-ME2 and PKA1 density functionals~\cite{Moreno-Torres2010PRC}, such difference extracted from nuclear masses in an effective way is around 2 MeV for $^{42}$Si.
Our additional DRHBc calculations using NL3* and PK1 density functionals show that the $1f_{7/2}$-$2p_{3/2}$ energy difference is smaller than 1.0 MeV for $^{40}$Mg and smaller than 2.5 MeV for $^{42}$Si in the spherical limit, and both have deformed ground states with $|\beta_2| > 0.3$, providing a consistent explanation for the suggested $N = 28$ shell quenching~\cite{Bastin2007PRL,Doornenbal2013PRL}.
Another $N = 28$ isotone, $^{41}$Al, may even exhibit triaxial deformation as predicted by the triaxial relativistic Hartree-Bogoliubov theory in continuum~\cite{Zhang2022arXiv}, and the $1f_{7/2}$-$2p_{3/2}$ energy difference in the spherical limit is around 1 MeV in the PC-PK1 result.
Thus, the proximity of $2p_{3/2}$ and $1f_{7/2}$ orbitals in the spherical limit might be common in the covariant density functional theory for $N = 28$ isotones from $^{39}$Na to $^{42}$Si, which are very neutron-rich.
Note that in stable nuclei the pseudospin symmetry would lead to the near-degeneracy of $2p_{3/2}$ and $1f_{5/2}$ orbitals~\cite{Liang2015PhysRep}.
In fact, the erosion of $N = 28$ shell closure near the neutron dripline for $10 \le Z \le14$ is also manifested in terms of the evolution of two-neutron separation energies and two-neutron gaps from the DRHBc mass table for even-even nuclei~\cite{Zhang2022ADNDT}.

In axially deformed cases, each $nlj$ orbital is split into $2j+1$ ones ($\frac{2j+1}{2}$ displayed in Fig.~\ref{fig2} because of Kramers' degeneracy) with quantum numbers $\Omega^\pi$, where $\pi$ is the parity and $\Omega$ the projection of angular momentum on the symmetry axis.
The pairing correlations and the $pf$ component mixing driven by deformation lead to the partial occupation of the $1/2^-$ and $3/2^-$ orbitals with certain $p$-wave components.
The valance neutrons with $s$- or $p$-wave nature in weakly bound nuclei can tunnel far out into the classically forbidden region, and a diffuse neutron density distribution or possibly a neutron halo occurs~\cite{Hansen1995ARNPS,Meng2006PPNP,Tanihata2013PPNP,Meng2015JPG}.
The lowered $2p$ states have been revealed to play a crucial role in the formation of halos close to the island of inversion, e.g., in $^{29}$F~\cite{Bagchi2020PRL,Casal2020PRC,Fossez2022PRC} and $^{31}$Ne~\cite{Nakamura2009PRL,Zhong2022SciChina}.
Given its small two-neutron separation energy, a two-neutron halo might be expected in $^{39}$Na, which was also indicated by the neutron density profiles obtained from the Skyrme-Hartree-Fock-Bogoliubov calculations~\cite{Chai2020PRC,Chai2022Symmetry}.

\begin{figure}[htbp]
  \centering
  \includegraphics[width=0.4\textwidth]{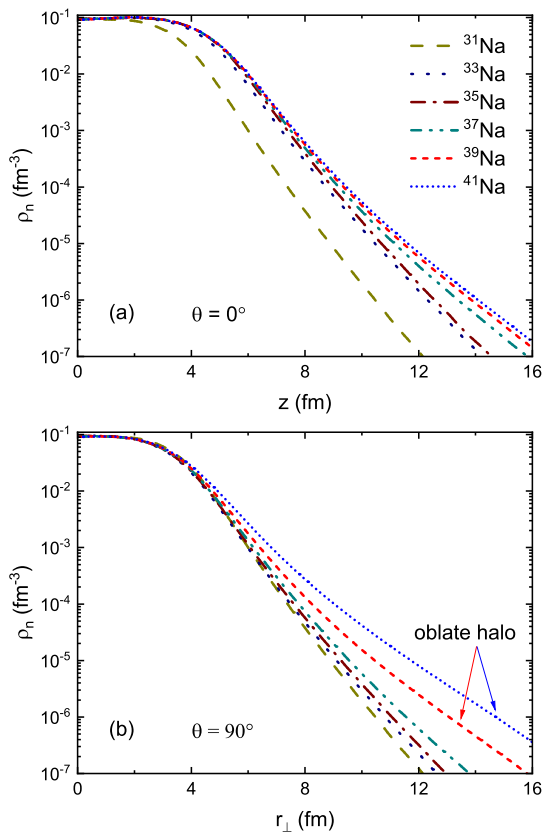}
  \caption{Neutron density distributions (a) along ($\theta = 0^\circ$) and (b) perpendicular to ($\theta = 90^\circ$) the symmetry axis for neutron-rich odd-even sodium isotopes $^{31,33,\cdots,41}$Na. In (b), $r_\perp = \sqrt{x^2+y^2}$.}
\label{fig3}
\end{figure}

To examine further the possible halo in $^{39}$Na, the neutron density distributions along and perpendicular to the symmetry axis for neutron-rich odd-even sodium isotopes with $N\ge20$ are shown in Fig.~\ref{fig3}.
The ground-state deformation of $^{31}$Na is spherical due to the $N=20$ shell closure, and $^{33,35,37}$Na are well deformed with quadrupole deformation $\beta_2 > 0.35$.
Therefore, the significant increase in density distribution along the symmetry axis from $^{31}$Na to others shown in Fig.~\ref{fig3}(a) can be understood from the deformation effects.
From $^{33}$Na to $^{41}$Na, the neutron density distribution along the symmetry axis gradually becomes more diffuse with the increasing neutron number.
In Fig.~\ref{fig3}(b), similar gradual growth is seen in the neutron density distribution perpendicular to the symmetry axis from $^{31}$Na to $^{37}$Na, while remarkable increases at large $r_\perp$ are found for $^{39}$Na and $^{41}$Na, even though they are prolately deformed.
This suggests that an oblate neutron halo that is mainly distributed perpendicular to the symmetry axis is formed around the prolate core in $^{39,41}$Na.

\begin{figure}[htbp]
  \centering
  \includegraphics[width=0.35\textwidth]{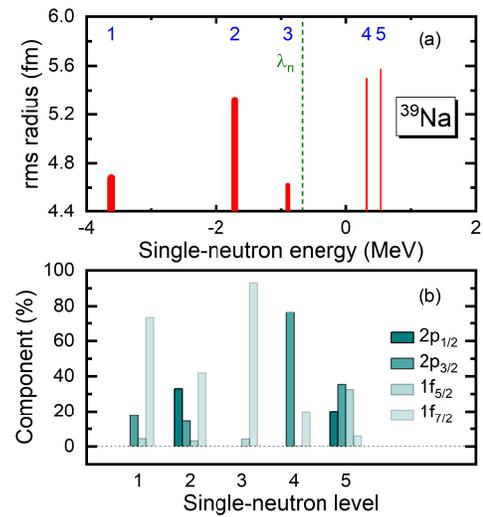}
  \caption{(a) The rms radius versus the energy $\epsilon$ for single-neutron levels around the Fermi energy $\lambda_n$ in the canonical
                  basis for $^{39}$Na. The thickness of each level is proportional to its occupation probability. (b) The main spherical components of the single-neutron levels labelled by numbers 1-5 in (a).}
\label{fig4}
\end{figure}

To understand the formation of the oblate halo, the root-mean-square (rms) radii and main spherical components of the single-neutron levels around the Fermi energy are shown in Fig.~\ref{fig4} for $^{39}$Na.
The levels are labelled by numbers according to the single-neutron energies in Fig.~\ref{fig4}(a), and their quantum numbers are respectively $3/2^-$, $1/2^-$, $5/2^-$, $1/2^-$, and $3/2^-$ as seen in Fig.~\ref{fig2}.
The rms radii of levels 4 and 5 that are embedded in the continuum and the bound level 2 are notably larger than those of levels 1 and 3.
This can be understood from the spherical components shown in Fig.~\ref{fig4}(b).
The considerable components of $2p_{1/2}$ and $2p_{3/2}$ account for the large rms radii of levels 2, 4, and 5.
In contrast, the large centrifugal barrier of $f$-wave components hinders strongly the spatial extension of wave functions for levels 1 and 3.
Note that level 2 is relatively deep bound with an energy $\epsilon\approx -2$~MeV and there are already nearly two valance neutrons occupying the levels above the Fermi energy.
Thus level 2 would not contribute to the halo.

Next we analyze the shape decoupling phenomenon.
For level 4 with $\Omega^\pi = 1/2^-$, both spherical harmonic functions $|Y_{10}(\theta,\varphi)|^2$ and $|Y_{1\pm1}(\theta,\varphi)|^2$ from $2p$ states contribute according to the angular momentum coupling, and the latter dominates.
For level 5 with $\Omega^\pi = 3/2^-$, only $|Y_{1\pm1}(\theta,\varphi)|^2$ contributes.
The angular distribution of $|Y_{10}(\theta,\varphi)|^2 \propto \cos^2\theta$ is prolate, while that of $|Y_{1\pm1}(\theta,\varphi)|^2 \propto \sin^2\theta$ is oblate.
Therefore, an oblate halo is expected for $^{39}$Na.
This also demonstrates that the deformation of a halo depends essentially on the quantum numbers of the halo orbitals and their main components.

\begin{figure}[htbp]
  \centering
  \includegraphics[width=0.5\textwidth]{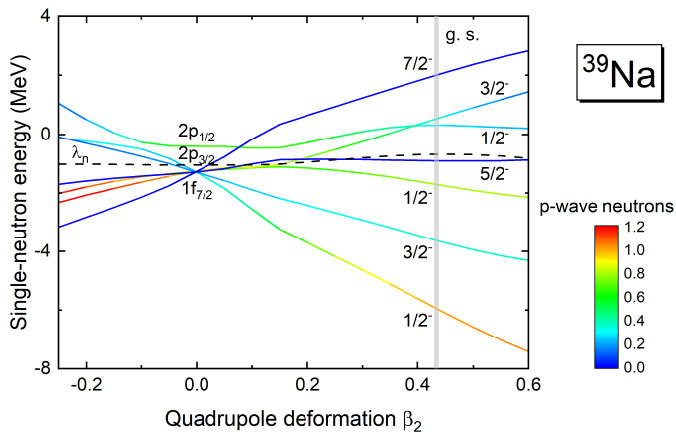}
  \caption{Same as Fig.~\ref{fig2} but only with $2p$ and $1f_{7/2}$ orbitals. Here the occupation number of $p$-wave neutrons is scaled by colors.}
\label{fig5}
\end{figure}

As shown in Fig.~\ref{fig3}, a more pronounced neutron halo might develop in $^{41}$Na.
This is because the increase of neutron numbers occupying the halo orbitals $1/2^-$ and $3/2^-$ (labelled by Nos. 4 and 5 in Fig.~\ref{fig4}) after adding two more neutrons;
although the considerable $2p$ components give rise to the formation of the halo in $^{39}$Na, the halo orbitals are only partially occupied as shown in Fig.~\ref{fig2}, resulting in a limited number of $p$-wave neutrons.
To see this intuitively, the number of $p$-wave neutrons in the single-neutron levels split from $2p$ and $1f_{7/2}$ orbitals are shown in Fig.~\ref{fig5}.
This number is calculated by $2v^2 \times \sum |C_p|^2$ for each level, where $v^2$ is its occupation probability and $C_p$s are the coefficients of its wave function on the $p$-wave Dirac Woods-Saxon bases.
For the ground state of $^{39}$Na, each halo orbital contributes less than 0.3 $p$-wave neutrons, summing to 0.56, about $30\%$ of the valance neutrons.
For $^{41}$Na, the ground state is less deformed compared to $^{39}$Na, which, together with the mean field change induced by adding neutrons, lowers the halo orbitals $1/2^-$ and $3/2^-$.
As a result, the two more neutrons in $^{41}$Na make the halo orbitals more occupied and contribute more $p$-wave neutrons.
It turns out that more than 1.2 $p$-wave neutrons are contributed, leading to a more prominent halo in $^{41}$Na.


In summary, the newly discovered $^{39}$Na with the neutron number $N=28$ is investigated within the deformed relativistic Hartree-Bogoliubov theory in continuum.
Based on several relativistic density functionals, $^{39}$Na is found to be well deformed in its ground state.
From the single-neutron levels around the Fermi energy, it is revealed that the lowering of $2p_{1/2}$ and $2p_{3/2}$ orbitals in the spherical limit results in the collapse of the $N=28$ shell closure in $^{39}$Na.
The pairing correlations and the $pf$ component mixing driven by deformation lead to the partial occupation of the $1/2^-$ and $3/2^-$ orbitals with certain $p$-wave components, giving rise to the formation of a neutron halo.
The density profiles of neutron-rich sodium isotopes suggest that an oblate halo is formed around the prolate core in $^{39,41}$Na, adding them as new candidates of deformed halo nuclei with shape decoupling.
A microscopic examination of the rms radii, the main components, and the neutron numbers occupying $p$-wave orbitals unravels the mechanisms behind the shape decoupling phenomenon and the development of halos towards the dripline.

The calculated $p$ orbital percentage of $\sim30\%$ for the valance neutrons in $^{39}$Na would be valuable for the spectroscopic factors obtained in future experiments.
In addition, the two-neutron halo in $^{39}$Na would belong to the category of ``Borromean halos".
This particular three-body dynamics would be relevant to its further study through different nuclear reactions, e.g., the electric dipole response of low-lying excitations.
The measurements of reaction cross sections, core momentum distributions, and Coulomb dissociations would also be helpful in uncovering the mystery of $^{39}$Na.
Finally, the question of which observables are closely related to shape decoupling effects in deformed halos remains unanswered.
It is anticipated that future precise measurements on the density distribution and scattering of hadronic probes may reveal signals that shed light on this novel phenomenon.

\begin{acknowledgments}
Fruitful discussions with members of the DRHBc Mass Table Collaboration are gratefully appreciated.
This work was partly supported by the National Natural Science Foundation of China (Grant Nos.~U2230207, U2030209, 11935003, 11875075, 11975031, 12141501, 12070131001, and 12205308),
the National Key R\&D Program of China (Grant Nos.~2020YFA0406001, 2020YFA0406002, and 2018YFA0404400),
the China Postdoctoral Science Foundation (Grant No.~2022M713106),
and the Institute for Basic Science (IBS-R031-D1).
P. P. was supported by the Rare Isotope Science Project of Institute for Basic Science, funded by the Ministry of Science and ICT (MSICT), and National Research Foundation of Korea (2013M7A1A1075764).
M.-H. M. was supported by the National Research Foundation of Korea (NRF) grants funded by the Korean government (Ministry of Science and ICT) (Grant Nos.~NRF-2021R1F1A1060066, NRF-2020R1A2C3006177, and NRF-2021R1A6A1A03043957).
The results described in this paper were obtained using High-performance computing Platform of Peking University and the High-Performance Computing Cluster of the Institute of Theoretical Physics, Chinese Academy of Sciences.
A portion of the computational resources were provided by the National Supercomputing Center including technical support (No.~KSC-2022-CRE-0333).
\end{acknowledgments}

%

\end{document}